\documentclass[12pt]{article}
\usepackage{cite}
\usepackage{amsbsy}

\title{Test Particles with Acceleration-Dependent Lagrangian.}
\author{M. Toller \thanks{e-mail: toller@iol.it}\\ 
via Malfatti n. 8  \\
I-38100 Trento, Italy}
 
\begin{document} 
\maketitle
                 
\begin{abstract}
We consider a classical test particle subject to electromagnetic and gravitational fields, described by a Lagrangian depending on the acceleration and on a fundamental length.  We associate to the particle a moving local reference frame and we study its trajectory in the principal fibre bundle of all the Lorentz frames. We discuss in this framework the general form of the Lagrange equations and the connection between symmetries and conservation laws (Noether theorem).  We apply these results to a model, already discussed by other authors, which implies an upper bound to the proper acceleration and to another new model in which a similar quantity, called ``pseudo-acceleration'', is bounded.  With some simple choices of the fields, we illustrate some other interesting properties of the models and we show that unwanted features may appear, as unstable run-away solutions and unphysical values of the energy-momentum or of the velocity.
\bigskip

 
45.20.Jj (Lagrangian and Hamiltonian mechanics); 

11.30.Cp (Lorentz and Poincar\'e invariance in particles and fields).  
\end{abstract}

\newpage

\section{Introduction.}

At present, the only well established long range relativistic fields, that allow a classical (non quantum) treatment, are the Maxwell electromagnetic field and the Einstein gravitational field. However, the existence of new long range fields, or of modifications of the above mentioned ones is often suggested on the basis of theoretical speculations or of not yet well confirmed experimental results. Many crucial problems of present day physics and cosmology can be formulated in terms of long range fields, for instance a possible failure of the Lorentz symmetry can be attributed to a long range nonscalar field. 

A long range field can be defined in terms of well chosen test particles. For instance, charged particles are needed to define the electromagnetic field and only spinning particles are affected by a torsion field \cite{HHKN}. We see that the investigations of long range classical fields and of test particles are intimately related. The aim of the present article is to give a contribution to this problem by discussing new kinds of test particles.

By definition, in a test particle model one disregards the influence of the particle on the field, namely several important effects, as the radiated energy, the radiation damping, the electromagnetic contribution to the mass, etc.  However, these models are much simpler and permit a better undertanding of some general features of the theory, in particular its symmetry properties. 

The test particle models may also suggest a way to introduce in the theory a fundamental length $\lambda$ of the order of the Planck length with the aim of taking into account approximately some effect of quantum gravity. For instance, as it has been proposed by several authors \cite{Caianiello,CDFMV,Brandt1,Scarpetta,Brandt,FLPS,Schuller,SP,Papini}, an effect of quantum gravity should be an upper bound of the order of $\lambda^{-1}$ to the proper acceleration of particles (we use the convention $c = 1$). Note that the gravitational acceleration vanishes in a local inertial frame and a discussion of the maximal acceleration principle requires the presence of an electromagnetic field. More references are given in the report \cite{Toller1}, dedicated to a discussion and a comparison of various geometric descriptions of the maximal acceleration principle. Anther effect of the fundamental length is the appearance of new degrees of freedom and of excited states of the particle.

In order to study the motion of a test particle, it is useful to associate to every point of its world line, parametrized by the variable $\tau$, a moving local Lorentz frame (a tetrad) $s(\tau)$ and to consider the motion of the point $s(\tau)$ in the principal bundle $\mathcal{S}$ of all the local Lorentz frames. This point of view has been adopted in refs.\ \cite{Kunzle,Vanzo} in the framework of presymplectic dynamics \cite{Souriau} and in refs.\ \cite{SZ,Toller2,TV} starting from the balance equations of a field theory and following the ideas of ref.\ \cite{Dixon}. 

The method of moving rest frames has been applied in refs.\  \cite{Nesterenko2,Nesterenko,NFS,NFS2} to Lagrangians depending on the acceleration and also on higher derivatives of the velocity. In the following we develop a more general formalism which is not restricted to rest moving frames.

The geometry based on the space $\mathcal{S}$ of the local Lorentz frames permits a formulation of a wide class of (possibly nonlocal) field theories \cite{CSTVZ} and particle theories. In the present article we concentrate our attention on the innovative particle models,  without a specific consideration of fields different from classical electromagnetism and  Einstein's gravitation. The general formalism, however, is ready for the introduction of more general classical fields. It has been remarked \cite{Toller3,Toller4} that this geometry permits an elegant treatment of the upper bounds to the acceleration  and the angular velocity of the frames and of the (possibly extended) associated particles.

An important issue is the description of the motion of a maximally accelerated particle in a Lagrangian or Hamiltonian formalism, since it is necessary for the construction of a quantum theory. It has been treated in refs.\ \cite{NFS,NFS2}, by means of the idea, already mentioned above, of associating to the particle a moving local Lorentz frame. As it looks natural, rest frames of the particle were chosen, but we shall see in the following that this choice is too restrictive.  

The report \cite{Toller5} contains a treatment of various kinds of test particles by means of Lagrangians that depend on the acceleration and on the angular velocity and, for dimensional reasons, must contain a  parameter $\lambda$ that can be interpreted as a fundamental length. This report also contains a treatment of a pre-symplectic formalism, which is more powerful for the treatment of some models. In the present article we concentrate our attention on Lagrangians not depending on the angular velocity of the moving frame. 

In section II we summarize the relevant aspects of the geometry of the fibre bundle $\mathcal{S}$ and of its generalizations. In section III we develop a general Lagrangian dynamics for a point moving in this space. In section IV we introduce the additional assumptions concerning a test particle subject to an Einstein-Maxwell field. In section V we use the Noether theorem to find the conserved quantities corresponding to various kinds of symmetry properties of the system. 

In section VI we give a detailed discussion of the relation between the particle and the local reference frames associated to it. We show that, if the Lagrangian depends on the acceleration, the center of mass cannot coincide with the origin of the local frame, where the electric charge is assumed to be concentrated. In this situation, the concept of acceleration is ambiguous and, besides the proper acceleration of the point charge, we define a new different quantity, called pseudo-acceleration.

As a first example, in section VII we give two equivalent treatments of the ordinary spinless charged masssive particle.  In section VIII we apply the general methods developed in the preceding sections to the model treated in refs.\  \cite{NFS,NFS2}, based on rest frames. The model has upper bounds to the velocity and to the proper acceleration of the charge, but the energy-momentum four-vector may be spacelike. Moreover, even in the absence of external fields, the solutions present an unstable behaviour (run-away solutions) when the initial conditions are slightly modified. 

In section IX we introduce a different model based on zero-momentum frames. It has the correct energy-momentum spectrum and an upper bound, which does not concern the acceleration, but the pseudo-acceleration, requiring a slight reformulation of the maximal acceleration principle. 

In sections X and XI we study further this new model by considering a particle in a constant electromagnetic field and we show that the solutions have good stability properties. In some circumstances, the velocity of the charge may exceed the velocity light, but, since it rotates with a very large frequency, this effect is hardly observable.  In section XII we consider a particle in a simple curved space and, even if the curvature is small, we find a serious form of instability.

In section XIII we analyze another model, with a physical energy-momentum spectrum, and the charge moving slower than light. However, the acceleration and the pseudo-acceleration have no upper bounds. In section XIV we draw some general conclusions and we suggest some remarks on the quantization.

\section{The bundle of Lorentz frames.}

In the absence of an electromagnetic field, a powerful method to treat the dynamics of a (possibly extended) particle is to consider, for each value of the parameter $\tau$, a local Lorentz frame (tetrad) $s(\tau)$ in such a way that a ``distinguished point'' of the particle (not necessarily the center of mass) coincides with the origin of the frame (sometimes called a {\it moving frame}). We indicate by $\mathcal{S}$ the ten-dimensional bundle of the Lorentz frames of the space-time and we describe the motion of the particle by means of the line $\tau \to s(\tau) \in \mathcal{S}$.

Particle dynamics in this space has been introduced by K\"unzle \cite{Kunzle} (see also \cite{Vanzo}) in the framework of the presymplectic formalism \cite{Souriau}. A treatment based on the balance equations for the densities of energy, momentum and relativistic angular momentum, strongly influenced by Dixon's work \cite{Dixon}, is given in refs.\ \cite{SZ,Toller2,TV}. In the next section we develop a geometric Lagrangian formalism which  avoids the introduction of coordinate systems in the space $\mathcal{S}$. It is more general than the Lagrangian formalism used in ref.\ \cite {NFS}, because $s(\tau)$ is not assumed to be a rest frame.

The elements of $\mathcal{S}$  are orthonormal tetrads $\{e_0,\ldots, e_3\}$ of four-vectors in the pseudo-Riemannian space-time manifold $\mathcal{M}$. We assume that $\mathcal{M}$ is time-oriented and $e_0$ belongs to the future cone. We consider ten vector fields $A_0,\ldots, A_9$ in the manifold $\mathcal{S}$.  The fields  $A_0,\ldots, A_3$ generate parallel displacements of the tetrads along the directions of the tetrad vectors, $A_4 = A_{[23]}$, $A_5 = A_{[31]}$, $A_6 = A_{[12]}$ generate rotations around the spatial vectors of the tetrad and $A_7 = A_{[10]}$, $A_8 = A_{[20]}$, $A_9 = A_{[30]}$ generate Lorentz boosts along the same spatial vectors. The latin indices $i, k, j, l, m, n$ take the values $0,\ldots, 3$ and the greek indices $\alpha, \beta, \gamma$ take the values $0,\ldots, 9$. We assume $A_{[ik]} = A_{[ki]}$ and, when necessary, we use the square brackets to indicate that an antisymmetric pair of latin indices stands for a greek index. 

The vector fields $A_{\alpha}$ can also be considered as first order differential operators and their commutators (Lie brackets) can be written in the form
\begin{equation} \label{Coeff}
[A_{\alpha}, A_{\beta}] = F_{\alpha \beta}^{\gamma}  A_{\gamma}.
\end{equation}
The quantities $F_{\alpha \beta}^{\gamma} = - F_{\beta \alpha}^{\gamma}$, are called {\it structure coefficients} and in the absence of gravitation they are the structure constants of the Poincar\'e group. In the presence of gravitation, $F_{ik}^{[jl]}$ are the components of the curvature tensor and $F_{ik}^j$ are the components of the torsion tensor, which vanishes in Einstein's theory.

The ten dimensional manifold $\mathcal{S}$ has a structure of principal fibre bundle \cite{KN} with base $\mathcal{M}$ and structural group $SO^{\uparrow}(1, 3)$, but the details of this structure are not needed in the next section. The only relevant feature is that $\mathcal{S}$ is a $n$-dimensional differentiable manifold in which $n$ differentiable vector fields $A_{\alpha}$ are defined and that, for each point $s \in \mathcal{S}$, the vectors $A_{\alpha}(s)$ form a basis of the tangent space $T_s\mathcal{S}$. As a consequence, we can identify all the tangent spaces with a single $n$-dimensional vector space $\mathcal{T}$, which, in the absence of gravitation and other external fields, is the Lie algebra of the Poincar\'e group. 

In the absence of gravitational field, $\mathcal{S}$ is the bundle of the Lorentz frames of the Minkowski space-time and the orthochronous Poincar\'e group $\mathcal{P}$ acts freely and transitively on $\mathcal{S}$. We choose a fixed frame $\hat s$ and represent univocally all the other frames in the form $s = g\hat s$ with $g \in \mathcal{P}$. In this way we may identify $\mathcal{S}$ and $\mathcal{P}$. The vector fields $A_{\alpha}$ generate the left translations of $\mathcal{P}$.

We shall also use the differential 1-forms $\omega^{\beta}$ defined by 
\begin{equation}
\omega^{\beta}(A_{\alpha}) = \delta_{\alpha}^{\beta}.
\end{equation}
The vectors $\omega^{\beta}(s)$ provide a basis in the cotangent space $T^*_s\mathcal{S}$. 

More details on this kind of geometry can be found in ref.\ \cite{CSTVZ}. In this way we can also treat a large class of external fields, including torsion fields \cite{HHKN}, and, for $n > 10$, gauge fields \cite{CSVZ}, including electromagnetic fields \cite{Toller}, as it is explained in section IV.

\section{The Lagrange equations.}

We write the tangent vectors of the curve $\tau \to s(\tau) \in \mathcal{S}$, which  describes the motion of the frame, in the form
\begin{equation} \label{Deriv}
\frac{ds(\tau)}{d \tau} = b^{\alpha} A_{\alpha}(s(\tau))
\in \mathcal{T}
\end{equation}
and we consider the action principle 
\begin{equation}
\delta \int_{\tau_1}^{\tau_2} L(b^{\alpha}, s) \, d \tau = 0.
\end{equation} 
where the Lagrangian $L$ is an homogeneous function of degree one of the ``velocities'' $b^{\alpha}$.

In order to derive the dynamical equations, we consider a family, parametrized by $\epsilon$, of varied trajectories $(\epsilon, \tau) \to s(\epsilon, \tau)$ with the usual properties
\begin{equation} \label{Conditions}
s(0, \tau) = s(\tau), \qquad
s(\epsilon, \tau_1) = s(\tau_1), \qquad s(\epsilon, \tau_2) = s(\tau_2).
\end{equation}
We put
\begin{equation}
\frac{\partial s(\epsilon, \tau)}{\partial \epsilon} = a^{\alpha} A_{\alpha}.
\end{equation}

If $\phi(s)$ is a differentiable auxiliary function, we have
\begin{equation}
\frac{\partial \phi(s)}{\partial \epsilon} = a^{\alpha} A_{\alpha}\phi, \qquad
\frac{\partial \phi(s)}{\partial \tau} = b^{\alpha} A_{\alpha}\phi,
\end{equation}
\begin{equation}
\frac{\partial^2 \phi(s)}{\partial \epsilon \partial \tau} = 
a^{\alpha} A_{\alpha} (b^{\beta} A_{\beta}\phi) =
b^{\alpha} A_{\alpha} (a^{\beta} A_{\beta}\phi).
\end{equation}
From the last equality we obtain
\begin{equation}
\left((a^{\alpha} A_{\alpha} b^{\beta}) A_{\beta} - 
(b^{\alpha} A_{\alpha} a^{\beta}) A_{\beta} +
a^{\alpha} b^{\beta} [A_{\alpha}, A_{\beta}] \right)\phi = 0,
\end{equation}
namely
\begin{equation}
\left(\frac{\partial b^{\gamma}}{\partial \epsilon} - 
\frac{\partial a^{\gamma}}{\partial \tau} +
a^{\alpha} b^{\beta} F_{\alpha \beta}^{\gamma} \right) A_{\gamma} \phi = 0,
\end{equation}
and, since $\phi$ is arbitrary,
\begin{equation}
\frac{\partial b^{\gamma}}{\partial \epsilon}  = 
\frac{\partial a^{\gamma}}{\partial \tau} -
a^{\alpha} b^{\beta} F_{\alpha \beta}^{\gamma}
\end{equation}
and finally (disregarding higher order terms in $\epsilon$) 
\begin{equation} \label{DeltaB}
\delta b^{\alpha} = 
\epsilon \left( \frac{\partial b^{\alpha}}{\partial \epsilon} \right)_{\epsilon = 0} = \epsilon \left( \frac{d a^{\alpha}}{d \tau} -
a^{\beta} b^{\gamma} F_{\beta \gamma}^{\alpha} \right)_{\epsilon = 0}
= \epsilon b^{\gamma} \left( A_{\gamma} a^{\alpha} -
a^{\beta} F_{\beta \gamma}^{\alpha} \right)_{\epsilon = 0}.
\end{equation}

By means of the last formula, one can write, performing a partial integration,
\begin{eqnarray} \label{Variation}
&\delta \int_{\tau_1}^{\tau_2} L \, d \tau =
\epsilon \int_{\tau_1}^{\tau_2} \left(\frac{\partial L}{\partial b^{\alpha}} 
\left( \frac{d a^{\alpha}}{d \tau} -
a^{\beta} b^{\gamma} F_{\beta \gamma}^{\alpha} \right) 
+ a^{\alpha} A_{\alpha} L \right)\, d \tau =& \nonumber \\
&\epsilon \int_{\tau_1}^{\tau_2} \left(- \frac{d}{d \tau} \frac{\partial L}{\partial b^{\alpha}} 
- \frac{\partial L}{\partial b^{\beta}}
b^{\gamma} F_{\alpha \gamma}^{\beta} +  A_{\alpha} L \right) a^{\alpha} \, d \tau + 
\epsilon \left[\frac{\partial L}{\partial b^{\alpha}} a^{\alpha} \right]_{\tau_1}^{\tau_2}.&
\end{eqnarray} 
The last term vanishes as a consequence of the conditions (\ref{Conditions}) and, considering that $a^{\alpha}$ is an arbitrary function of $\tau$, we obtain the Euler-Lagrange dynamical equations
\begin{equation} \label{Dyn3}
\dot p_{\alpha} = \frac{d p_{\alpha}}{d \tau} =
b^{\gamma} p_{\beta} F_{\gamma \alpha}^{\beta} - A_{\alpha} L,
\end{equation} 
where
\begin{equation} \label{Dyn4}
p_{\alpha} = - \frac{\partial L}{\partial b^{\alpha}}. 
\end{equation} 

If we assume that $L$ does not depend directly on $s$, the last term in eq.\ (\ref{Dyn3}) is not present and this equation is exactly the one obtained in ref.\ \cite{Toller2} in the pole approximation by integrating the balance equations (the quantities $F_{\alpha \beta}^{\gamma}$ are defined there with a different sign). 

The momenta $p_{\alpha}$ defined by eq.\ (\ref{Dyn4}) are homogeneous functions of degree zero of the velocities $b^{\alpha}$, namely they depend only on the ratios $(b^0)^{-1} b^{\alpha}$ (and possibly on $s$). It follows that they must satisfy at least a primary constraint \cite{Dirac}.   Other primary constraints may exist and we write them in the form
\begin{equation} \label{Constraints}
\Phi_{\rho}(p_{\alpha}, s) = 0, \qquad \rho = 0,\ldots, m - 1.
\end{equation} 

Since $L$ is an homogeneous function, the Euler theorem gives, taking eq.\ (\ref{Dyn4}) into account,
\begin{equation}
L + p_{\alpha} b^{\alpha} = 0
\end{equation}
and by differentiation we obtain
\begin{equation} \label{Diff}
b^{\alpha} dp_{\alpha} + (A_{\alpha} L) \, \omega^{\alpha}  = 0.
\end{equation}
The differentials $dp_{\alpha}$ and the forms $\omega^{\alpha}$ are arbitrary, apart from the constraints 
\begin{equation} \label{DiffConstr} 
\frac{\partial \Phi_{\rho}}{\partial p_{\alpha}} dp_{\alpha} + 
(A_{\alpha} \Phi_{\rho}) \, \omega^{\alpha} = 0, \qquad 
\rho = 0,\ldots, m - 1
\end{equation}
and from eq.\ (\ref{Diff}) we have
\begin{equation}  \label{Dyn5}
b^{\alpha} = \sum_{\rho} \alpha^{\rho} \frac{\partial \Phi_{\rho}}{\partial p_{\alpha}}, 
\end{equation}
\begin{equation}  \label{Deri}
A_{\alpha} L = \sum_{\rho} \alpha^{\rho} A_{\alpha}\Phi_{\rho}. 
\end{equation}

Eq.\ (\ref{Dyn5}) inverts, as far as possible, eq.\ (\ref{Dyn4}). At this level, the functions $\alpha^{\rho}(\tau)$ are arbitrary and, if they are not determined by the dynamical equations, they parametrize the gauge transformations of the system. There is at least one kind of gauge transformations, namely a redefinition of the parameter $\tau$.

If, according to the dynamical equations (\ref{Dyn3}), the quantities $\Phi_{\rho}$ are not conserved, the conditions $\dot \Phi_{\rho} = 0$ determine partially the functions $\alpha^{\rho}$ or give rise to secondary constraints. Some simplification can be obtained from the following consequence of the equations derived above
\begin{eqnarray} \label{Conserv}
&\sum_{\rho} \alpha^{\rho} \dot \Phi_{\rho} = 
\sum_{\rho} \alpha^{\rho} \left( \frac{\partial \Phi_{\rho}}{\partial p_{\alpha}} \dot p_{\alpha}  + A_{\alpha}\Phi_{\rho} b^{\alpha}\right) =& 
\nonumber \\  
&b^{\alpha} (\dot p_{\alpha} + A_{\alpha} L) = 
b^{\alpha} b^{\beta} p_{\gamma} F_{\beta \alpha}^{\gamma} = 0.&
\end{eqnarray}
It follows that if $\alpha^0 \neq 0$ and the secondary constraints $\dot \Phi_{\rho} = 0$ for $\rho = 1,\ldots, m-1$ are satisfied, $\Phi_0$ is conserved. In particular, if there is only one primary constraint, it is conserved and there are no secondary constraints.

If one is able to express all the quantities $b^{\alpha}$ and $\dot p_{\alpha}$ as smooth functions of of $s$ and $p_{\alpha}$ (satisfying some constraints), we say that the equations of motion are in {\it normal form} and one can apply the theorems on the (local) existence and uniqueness of the solutions. In fact, if one introduces a local system of coordinates $q^{\alpha}$ in the space $\mathcal{S}$, it follows from eq.\ (\ref{Deriv}) that the quantities $\dot q^{\alpha}$  are smooth functions of the quantities $b^{\alpha}$.

\section{Einstein-Maxwell fields.}

Now we consider a test particle moving in an Einstein gravitational field and a Maxwell electromagnetic field. In order to describe the electromagnetic field, we adopt the procedure indicated in refs.\ \cite{CSVZ,Toller2,Toller}, namely we introduce a principal fibre bundle $\mathcal{S}$ with base $\mathcal{M}$ and structural group $SO^{\uparrow}(1, 3) \times U(1)$ which includes the electromagnetic gauge group. Then the manifold $\mathcal{S}$ has dimension $n = 11$ and we have to introduce a new vector field, that we indicate by $A_{\bullet}$ (in order to avoid a two-digit index $\alpha = 10$), which generates the global electromagnetic gauge transformations. We use the notation
\begin{equation} \label{Deriv1}
b^{\alpha} A_{\alpha} = 
b^{i} A_{i} + \frac 1 2 b^{[ik]} A_{[ik]} + b^{\bullet} A_{\bullet},
\qquad b^{[ik]} = - b^{[ki]}.
\end{equation}
The eleven-dimensional manifold $\mathcal{S}$ can also be considered as a principal fibre bundle with structural group $U(1)$ and base $\mathcal{S}_0$, the ten-dimensional bundle of the Lorentz frames considered up to now. We shall consider later the projection $\mathcal{S} \to \mathcal{S}_0$ along the fibers generated by the vector  field $A_{\bullet}$.

The structure coefficients $F_{[ik] \beta}^{\alpha}$ coincide with the structure constants of the Poincar\'e algebra. They can be written in the form
\begin{eqnarray} \label{Structure}
&F_{[ik][jl]}^{[mn]} = \delta_i^m g_{kj} \delta_l^n - \delta_k^m g_{ij} \delta_l^n
- \delta_i^m g_{kl} \delta_j^n + \delta_k^m g_{il} \delta_j^n& \nonumber \\
&- \delta_i^n g_{kj} \delta_l^m + \delta_k^n g_{ij} \delta_l^m
+ \delta_i^n g_{kl} \delta_j^m - \delta_k^n g_{il} \delta_j^m,&
\end{eqnarray}
\begin{equation} \label{Structure1}
F_{[ik] j}^l = \delta_i^l g_{kj} - \delta_k^l g_{ij},
\end{equation}
\begin{equation} \label{Structure2}
F_{[ik][jl]}^m = 0, \qquad F_{[ik] j}^{[mn]} = 0.  
\end{equation} 
We also assume that the torsion $F_{ik}^j$ vanishes.

The structure coefficients $F_{ik}^{\bullet} = F_{ik}$ represent the electromagnetic field strength and $F_{\alpha \bullet}^{\beta} = F_{[ik] \alpha}^{\bullet} = 0$. The electromagnetic interaction Lagrangian is 
\begin{equation}
L_I = e b^{\bullet}, 
\end{equation} 
where $e$ is the electric charge.

By means of these equations, we can write eq.\ (\ref{Dyn3}) in the more explicit form
\begin{equation} \label{Dyn6}
\dot p_i = - b_i{}^k p_k + F_i, \quad 
F_i =  b^k G_{ik} - A_i L, \quad
G_{ik} = \frac 1 2 p_{[jl]} F_{ki}^{[jl]} + e F_{ik},
\end{equation}
\begin{equation} \label{Dyn7}
\dot p_{[ik]} = b_i p_k - b_k p_i - b_i{}^j p_{jk} - b_k{}^j p_{ij} - A_{[ik]} L,
\end{equation}
\begin{equation} \label{Carica}
\dot p_{\bullet} = 0, \qquad p_{\bullet} = - e,
\end{equation}
where $F_{ki}^{[jl]}$ is the Riemann curvature tensor which describes the gravitational field. As we discuss in the next section, the quantities $p^i$ and $-p_{ik}$ are interpreted as the energy, the momentum and the relativistic angular momentum of the particle, measured in the local reference frame $s$. The quantity $-p_{\bullet}$ is the conserved electric charge.

It is often convenient to use a three-dimensional vector formalism. We introduce the vectors 
\begin{equation} \label{Vector1}
\mathbf{b} =(b^1, b^2, b^3), \qquad
\mathbf{b'} =(b^{[23]}, b^{[31]}, b^{[12]}), \qquad
\mathbf{b''} =(b^{[10]}, b^{[20]}, b^{[30]}),
\end{equation}
\begin{equation} \label{Vector2}
\mathbf{p} = - (p_1, p_2, p_3), \quad
\mathbf{p'} = - (p_{[23]}, p_{[31]}, p_{[12]}), \quad
\mathbf{p''} = - (p_{[10]}, p_{[20]}, p_{[30]}),
\end{equation}
\begin{equation} 
\mathbf{f} = - (b^0)^{-1} (F_1, F_2, F_3),
\end{equation}
\begin{equation}
\mathbf{E} = (F_{01}, F_{02}, F_{03}), \qquad 
\mathbf{B} = (F_{32}, F_{13}, F_{21}),
\end{equation}
\begin{equation} \label{EB}
\hat \mathbf{E} = (G_{01}, G_{02}, G_{03}), \qquad 
\hat \mathbf{B} = (G_{32}, G_{13}, G_{21}),
\end{equation}
and we obtain, if $L$ does not depend on $s$,
\begin{equation} \label{PDot1}
\dot p_0 = - \mathbf{b''} \cdot \mathbf{p} + \mathbf{b} \cdot \mathbf{f},
\end{equation}
\begin{equation}  \label{PDot2}
\dot \mathbf{p} = - \mathbf{b'} \times \mathbf{p} - p_0 \mathbf{b''} + b^0 \mathbf{f},
\end{equation}
\begin{equation} \label{PDot3}
\dot \mathbf{p}' = - \mathbf{b} \times \mathbf{p}
- \mathbf{b'} \times \mathbf{p'} - \mathbf{b''} \times \mathbf{p''},
\end{equation} 
\begin{equation} \label{PDot4}
\dot \mathbf{p}'' = p_0 \mathbf{b} - b^0 \mathbf{p}
- \mathbf{b'} \times \mathbf{p''} + \mathbf{b''} \times \mathbf{p'},
\end{equation}
\begin{equation} \label{EBf}
\mathbf{f} = \hat \mathbf{E} + (b^0)^{-1} \mathbf{b} \times \hat \mathbf{B}.
\end{equation}

The dimension of the phase space, namely the number of parameters necessary to define the initial conditions, is given by $2n = 22$ minus the number of primary and secondary constraints (including the constraint (\ref{Carica})) minus the number of the arbitrary gauge parameters (including the usual electromagnetic gauge transformations).

It is instructive to consider, besides the flat Minkowski space-time, the spaces of constant curvature
\begin{equation} \label{CCurv}
F_{i k}^{[j l]} = \rho (\delta_i^j \delta_k^l - \delta_k^j \delta_i^l).
\end{equation}
$\mathcal{M}$ is a de Sitter spacetime if $\rho > 0$, and an anti de Sitter spacetime for $\rho < 0$. From eq.\ (\ref{Dyn6}) we obtain
\begin{equation}
G_{ik} =  - \rho p_{[ik]} + e F_{ik},
\end{equation}
or, in the vector notation,
\begin{equation} \label{CCurv2}
\hat \mathbf{E} = - \rho \mathbf{p''} + e \mathbf{E}, \qquad
\hat \mathbf{B} = - \rho \mathbf{p'} + e \mathbf{B}.
\end{equation}

\section{Noether's theorem.}

In order to treat the connection between symmetries and conservation laws (Noether's theorem), we consider a vector field
\begin{equation} \label{B1}
Y(s) = a^{\alpha}(s) A_{\alpha}(s)
\end{equation}
and the corresponding one-parameter diffeomorphism group $\exp(\epsilon Y)$, which transforms the trajectory $s(\tau)$ into the trajectories $s(\epsilon, \tau)$. If this transformation does not change the action, the expression (\ref{Variation}) vanishes and, taking the dynamical equations into account, we obtain the conservation law 
\begin{equation} \label{Noether}
a^{\alpha} p_{\alpha} = \rm{constant}.
\end{equation} 

We assume that $L$ does not depend on $s$ and consider two kinds of applications of this general theorem, which use two different kinds of symmetry properties. In the first case we require
\begin{equation} \label{B2}
[Y, A_{\beta}] = \left( a^{\alpha} F_{\alpha \beta}^{\gamma} - 
A_{\beta} a^{\gamma} \right) A_{\gamma} = 0
\end{equation}
and from eq.\ (\ref{DeltaB}) we see that $\delta b^{\alpha} = 0$, and therefore $\delta L = 0$.  Note that the validity of the conservation law (\ref{Noether}) does not depend on the form of the function $L(b^{\alpha})$.

In a second case we assume eqs.\ (\ref{Structure}--\ref{Structure2}) and consider the infinitesimal Lorentz transformation generated by $A_{[ik]}$. From eq.\ (\ref{DeltaB}) we obtain 
\begin{equation}
\delta b^{\alpha} = - \epsilon F_{[ik] \gamma}^{\alpha} b^{\gamma},
\end{equation}
namely an infinitesimal Lorentz transformation of the quantities $b^{\alpha}$. If the Lagrangian is invariant under this transformation, the quantity $p_{[ik]}$ is conserved. In particular, if the Lagrangian is a Lorentz scalar function of $b^{\alpha}$ all the six quantities $p_{[ik]}$ are conserved. If the Lagrangian is only a rotational scalar, only the three quantities $p_{[rs]}$ with $r, s = 1, 2, 3$ are conserved.  These conservation laws depend on the invariance properties of the Lagrangian and on the form of the structure coefficients $F_{[ik] \gamma}^{\alpha}$. They are also valid in the presence of arbitrary gravitational and electromagnetic fields.

Now we consider the simplest application of the first kind. As we have seen in section II, in the absence of gravitational and electromagnetic fields, $\mathcal{S}$ is the bundle of the Lorentz frames of the Minkowski space-time and choosing a fixed frame $\hat s$, can be identified with the orthochronous Poincar\'e group $\mathcal{P}$. The vector fields $A_{\alpha}$ generate the left translations of $\mathcal{P}$, but one can also introduce the vector fields $\hat A_{\alpha}$, which generate the right translations, interpreted as Poincar\'e transformations of the fixed frame $\hat s$. They commute with $A_{\alpha}$ and are given by
\begin{equation} 
Y = \hat A_{\alpha}(g) = D^{\beta}{}_{\alpha}(g) A_{\beta}(g),
\end{equation}
where $D(g)$ is the adjoint representation of $\mathcal{P}$, which has the property 
\begin{equation} \label{Adj}
A_{\alpha} D^{\beta}{}_{\gamma}(g) =  - F_{\alpha \delta}^{\beta} D^{\delta}{}_{\gamma}(g).
\end{equation} 

It follows from the Noether theorem that the quantities
\begin{equation}
\hat p_{\alpha} = D^{\beta}{}_{\alpha}(g(\tau)) p_{\beta}(\tau)
\end{equation}
are conserved. Since they are defined starting from the symmetry under spacetime translations and infinitesimal Lorentz transformations of $\hat s$, they have to be interpreted as the components of the energy-momentum and the relativistic angular momentum measured in the fixed frame $\hat s$. 

We indicate an element of $\mathcal{P}$ by $(\Lambda, x) = (\Lambda, 0)(1, x)$, where $x$ is the translation four-vector and $\Lambda$ is the Lorentz matrix acting on the contravariant components of the four-vectors. If we use the explicit form of the adjoint representation, we obtain the equations
\begin{equation} \label{Adj1}
\hat p_i = \Lambda^k{}_i  p_k, \qquad
\hat p_{[ik]} = \Lambda^j{}_i \Lambda^l{}_k p_{[jl]} - x_i \hat p_k + x_k \hat p_i,
\end{equation}
which show that the quantities $p_{\beta}(\tau)$ are correctly interpreted as as the components of the energy-momentum and the relativistic angular momentum measured in the moving frame $s(\tau)$, as we anticipated in the preceding section. The mass $\mu$ and the spin $\sigma $ of the particle are given by the familiar Poincar\'e invariant expressions and take the same form when written as functions of $\hat p_{\beta}(\tau)$ or of $p_{\beta}(\tau)$. In particular we have
\begin{equation} 
\mu^2 = p^i p_i, \qquad \sigma^2 \mu^2 = - S^k S_k, \qquad
S^k = 2^{-1} \epsilon^{kijl} p_{[ij]} p_l.
\end{equation}

This interpretation is also valid if an electromagnetic field is present. In fact, from eqs.\ (\ref{Vector1})-(\ref{EBf}) we see that the influence of the electromagnetic field on the derivatives  $\dot p_{\alpha}$ is just the one expected for a point charge centered at the origin, if the quantities $p_{\alpha}$ represent the components of the kinetic energy-momentum and relativistic angular momentum.  This problem is also discussed in ref.\ \cite{Nesterenko}. 

The exact definition of the energy-momentum and of the relativistic angular momentum of an extended particle in a curved spacetime is somehow ambiguous \cite{Dixon,Toller2} and in this case the interpretation of $p_{\alpha}$ can be considered as a definition. 

\section{Frames and particles.}

In order to interpret the solutions of the dynamical equations, we have to clarify the connection between the Lorentz frames $s(\tau)$ and the physical particle. We indicate by $\pi: \mathcal{S} \to \mathcal{M}$ the projection which associates to the tetrad $s \in \mathcal{S}$ its origin $x = \pi(s) \in \mathcal{M}$. The projection of the curve $\tau \to s(\tau)$ is the world line  $\tau \to \pi(s(\tau)) = x(\tau)$ in the the spacetime manifold $\mathcal{M}$. From eq.\  (\ref{Deriv}), remembering that the vector fields $A_i$ generate the parallel displacements along the four-vectors $e_i(\tau)$ of the tetrad $s(\tau)$, we have
\begin{equation}
\dot x(\tau) = \frac{dx(\tau)}{d \tau} = b^i(\tau) e_i(\tau).
\end{equation}
If $\tau$ is the proper time, namely if $b_i  b^i = 1$, the quantities $b^i$ are the components of the four-velocity of the origin of $s(\tau)$, with respect to the same frame.

The tetrad $s(\tau + d\tau)$ differs from the parallel transported tetrad by an infinitesimal Lorentz transformation with parameters $b^{[ik]} d \tau$. This means that the covariant derivatives of the tetrad four-vectors are given by
\begin{equation}
\frac{D e_i(\tau)}{d \tau} = - b_i{}^k(\tau) e_k(\tau),
\end{equation}
and we obtain the formula
\begin{equation}
a(\tau) = \frac{D \dot x(\tau)}{d \tau} = \dot b^i  e_i -  b^i b_i{}^k e_k =
(\dot b^i + b^i{}_k b^k) e_i. 
\end{equation}

If $\tau$ is the proper time, $a$ is the covariant acceleration four-vector and its components in the moving frame $s(\tau)$ are given by
\begin{equation}
a^i = \dot b^i + b^i{}_k b^k. 
\end{equation}

A {\it rest frame} is defined by the condition $\mathbf{b} = 0$ and the four-velocity of the origin of the frame is equal to the tetrad vector $e_0$, which is timelike by definition. If $\tau$ is the proper time, we also have $b^0 = 1$, and $d \tau$ is a time measured in the rest frame. We have
\begin{equation}
a = \frac{D e_0}{d \tau} =  b^{[r0]} e_r, \qquad  a^0 = 0, \qquad \mathbf{a} = \mathbf{b''}.
\end{equation}
The vector $\mathbf{a}$ represents the acceleration of the origin of a rest frame measured in the same frame. If the parameter $\tau$ is arbitrary, we can write the more general formula
\begin{equation} \label{Acc}
\mathbf{a} = (b^0)^{-1} \mathbf{b''}.
\end{equation}

We can introduce more specific moving frames by requiring the stronger conditions
\begin{equation}
\mathbf{b} = 0, \qquad b^{[20]} = b^{[30]} = b^{[31]} = 0.
\end{equation}
In a Lagrangian model they can be imposed by introducing six Lagrangian multipliers. In this case $s(\tau)$ is a Frenet frame and the quantities $k_1 = b^{[01]}$, $k_2 = b^{[12]}$, $k_3 = b^{[23]}$ are the geometric invariants of the world line, which depend on the acceleration and its first and second derivatives \cite{Nesterenko2,NFS}. The Lagrangian depends on these invariants as in the models considered in ref.\ \cite{Nesterenko2}.

Now we consider frames defined by non purely geometric, but dynamical requirements. The condition $\mathbf{p} = 0$ defines a {\it zero-momentum frame}. It implies that the energy-momentum is a timelike four-vector. In the case of a free particle, the center of mass is at rest in this frame, but this is not true in general. In a zero-momentum frame $\mathbf{p'}$ represents by definition the spin of the particle. One can still consider the vector (\ref{Acc}) but it does not represent the acceleration of the origin $x(\tau)$ and not even the acceleration of the center of mass. We call it the {\it pseudo-acceleration}. From eq.\ (\ref{PDot2}) we have
\begin{equation} 
\mathbf{a} = p_0^{-1} \mathbf{f}.
\end{equation}
Note that $p_0$ in a zero-momentum frame is the invariant mass and this formula coincides with the Newton formula, valid for a point particle. If the particle is extended, this formula remains valid if we replace the acceleration by the pseudo-acceleration. This may be considered as the physical meaning of the pseudo-acceleration.

In order to describe the position of the center of mass, we have to introduce for each tetrad a system of coordinates in a suitable open set of $\mathcal{M}$, for instance a system of normal coordinates. In the absence of gravitation, namely in special relativity theory,  one can associate to every local frame $s$ a Lorentzian coordinate system in the flat Minkowski spacetime and the space coordinates of the center of mass at zero time in this frame can be written in the form \cite{Moller}
\begin{equation} \label{COM2}
\mathbf{y} = - p_0^{-1} \mathbf{p''}.
\end{equation}
The condition $\mathbf{p''} = 0$ means that the trajectory of the center of mass crosses the origin. A frame with this property is called a {\it central frame}. This definition can be extended to the case in which gravitation is present. From eq.\ (\ref{PDot4}) we see that, if the spin is zero, a zero-momentum central frame is also a rest frame.

The definition $\mathbf{p''} = 0$ is not Lorentz invariant and it is often replaced by the condition
\begin{equation} \label{Dixon}
p_{[ik]} p^k = 0,
\end{equation}
proposed by Dixon \cite{Dixon}. A frame with this property is called a {\it Dixon frame}. It can always be obtained from a central zero-momentum frame by means of a Lorentz transformation.

The models treated in the following sections concern extended particles which contain a point charge. In this case, instead of working with central or Dixon frames, it is convenient to assume that the charge lies at the origin of $s(\tau)$. More in general, if the charge is not pointlike, one can require that the electric dipole moment with repect to the frame $s(\tau)$ vanishes.

\section{An ordinary spinless particle.}

It is instructive to consider first some Lagrangians that do not contain $\mathbf{b'}$ and $\mathbf{b''}$ and describe a pointlike spinless particle with mass $m$ and charge $e$. The simplest Lagrangian of this kind is
\begin{equation} 
L = - m b^0 + e b^{\bullet}.
\end{equation}
Since it is linear in the ``velocities'' $b^{\alpha}$, all the momenta $p_{\alpha}$ are fixed by eleven primary constraints, namely we have $p_0 = m$, $p_{\bullet} = -e$ and the other momenta vanish. As a consequence, the spin vanishes and the center of mass coincides with the charge. From the dynamical equations (\ref{PDot1}-\ref{PDot4}) we obtain
\begin{equation}  
\mathbf{b} = 0, \qquad m \mathbf{a} = m (b^0)^{-1} \mathbf{b''} = \mathbf{f} = e \mathbf{E}.
\end{equation}

We see that the frames $s(\tau)$ are zero-momentum rest central frames and that the acceleration is given by the usual formula. The time evolution of the quantities $b^0$, $\mathbf{b'}$ and $b^{\bullet}$ is not determined by the dynamical equations, the model has gauge invariances described by five parameters and the phase space has dimension six. If we impose the gauge fixing condition  $\mathbf{b'} = 0$, the frame is Fermi-Walker transported and if we assume  $b^0 = 1$ the parameter $\tau$ is the usual relativistic proper time. If $\mathbf{E} = 0$, the frame is parallel transported and the world line of the particle is a geodesic. 

In view of the following developments, it is interesting to treat with more detail a charged particle in a constant electromagnetic field in the absence of gravitation. The field is invariant under spacetime translations, but it changes if a Lorentz transformation is applied to the frame. If $b^0 = 1$, we have
\begin{equation} \label{VarBE}
\dot \mathbf{E} = \mathbf{b''} \times \mathbf{B} - \mathbf{b'} \times \mathbf{E}, \qquad
\dot \mathbf{B} = - \mathbf{b''} \times \mathbf{E} - \mathbf{b'} \times \mathbf{B}.
\end{equation}
Of course, the Lorentz invariants
\begin{equation} \label{Invar}
I = \|\mathbf{E}\|^2 - \|\mathbf{B}\|^2, \qquad
J = \mathbf{E} \cdot \mathbf{B}
\end{equation}
are constant.

The derivatives (\ref{VarBE}) vanish if we choose the gauge fixing condition
\begin{equation}
\mathbf{b'} = - e m^{-1} \mathbf{B}.
\end{equation}
The components of the field and all the quantities $b^{\alpha}$ are constant and we say that this is a ``stationary'' solution.  It follws that the projection of the trajectory on the ten-dimansional space $\mathcal{S}_0$, identified with the Poincar\'e group $\mathcal{P}$, is given by a one-parameter subroup, namely
\begin{equation} \label{OPS}
g(\tau) = \exp(\tau b^{\alpha} \tilde A_{\alpha}) g(0),
\end{equation}
where $\tilde A_{\alpha}$ form a basis in the Lie algebra of $\mathcal{P}$ and $\exp(.)$ is the exponential mapping of the group \cite{Kirillov}. 

We also consider the Lorentz invariant Lagrangian
\begin{equation}
L = - m \left((b^0)^2 - \|\mathbf{b}\|^2 \right)^{1/2} + e b^{\bullet}.
\end{equation}
We obtain
\begin{equation} 
p_0 = m b^0 \left((b^0)^2 - \|\mathbf{b}\|^2 \right)^{-1/2}, \qquad
\mathbf{p} = m \left((b^0)^2 -  \|\mathbf{b}\|^2 \right)^{-1/2} \mathbf{b},
\end{equation}
\begin{equation} 
\mathbf{p'} = \mathbf{p''} = 0, \qquad p_{\bullet} = - e 
\end{equation}
and the constraint
\begin{equation}
p_0^2 - \|\mathbf{p}\|^2 = m^2.
\end{equation}

It is sufficient to consider the dynamical equations (\ref{PDot2}), since eqs.\ (\ref{PDot1}),
(\ref{PDot3}), and (\ref{PDot4}) are consequences of the other equations. We see that the quantities  $b^0$, $\mathbf{b'}$, $\mathbf{b''}$ and $b^{\bullet}$ are not determined and they are gauge parameters. In particular, the Lorentz transformations assume the character of gauge transformations.  We may fix the gauge partially by assuming that $\mathbf{p} = 0$ and we obtain exactly the equations of the preceding model. The dynamics of the particle is the same, but in the second model there is more freedom in the choice of the moving frame.

Alternatively, we may require $\mathbf{b'} =\mathbf{b''} = 0$, namely that the moving frame is parallel transported, and, also assuming $b^0 = 1$,  eq.\ (\ref{PDot2}) takes the familiar form
\begin{equation} 
\dot \mathbf{p} = \mathbf{f}.
\end{equation}

\section{A model with a maximal acceleration.}

For dimensional reasons, a Lagrangian which is an homogeneous function of degree one of the velocities $b^\alpha$ and contains a fundamental length $\lambda$ must contain, besides $b^0$ and $\mathbf{b}$, some of the quantities $\lambda\mathbf{b'}$ or $\lambda\mathbf{b''}$. In the following we consider Lagrangians that depend on $\mathbf{b''}$, but not on $\mathbf{b'}$, and we treat first a model with a maximal proper acceleration equivalent to the one treated in ref.\ \cite{NFS}. 

It is easy to obtain an upper limit for the quantity $(b^0)^{-1} \mathbf{b''}$, but, as we have seen in section VI, this quantity coincides with the proper acceleration only if $s(\tau)$ are rest frames, namely if $\mathbf{b} = 0$. This constraint is implicit in the formalism of ref.\ \cite{NFS}, but in the framework of the preceding sections we have to enforce it by means of the Lagrange multipliers $\boldsymbol{\eta}$. A simple Lagrangian of this kind, that does not contain $\mathbf{b'}$, is given by
\begin{equation} \label{Lagr6}
L = - m \left((b^0)^2 - \lambda^2 \|\mathbf{b''}\|^2 \right)^{1/2} 
+ \boldsymbol{\eta} \cdot \mathbf{b} + e b^{\bullet}.
\end{equation} 

We obtain the equations
\begin{equation} 
\mathbf{b} = 0, \qquad \mathbf{p} = \boldsymbol{\eta}, \qquad \mathbf{p'} = 0, 
\qquad p_{\bullet} = - e.
\end{equation} 
The first equality means that we are dealing with rest frames and the second shows that $\mathbf{p}$ can be chosen freely in the initial conditions, even in contrast with the usual requirements on the energy-momentum spectrum. The third formula does not imply that the spin vanishes, because the spin is defined as the angular momentum in a zero-momentum frame, that in general is not a rest frame. 

The other momenta are given by
\begin{eqnarray} \label{PP}
&p_0 = m b^0 \left((b^0)^2 - \lambda^2 \|\mathbf{b''}\|^2 \right)^{-1/2},& \nonumber \\
&\mathbf{p''} = m \lambda^2 \left((b^0)^2 - \lambda^2 \|\mathbf{b''}\|^2 
\right)^{-1/2} \mathbf{b''}&
\end{eqnarray}
and satisfy the scalar primary constraint
\begin{equation}  \label{Constr2}
p_0^2 - \lambda^{-2} \|\mathbf{p''}\|^2 = m^2,
\end{equation}
which, in agreement with eq.\ (\ref{Conserv}), is conserved and does not give rise to secondary constraints. We consider only solutions with $p_0 \geq m$ and, as a consequence, $b^0 >0$. From the preceding formulas one obtains the relations
\begin{equation} \label{BB}
\mathbf{a} = (b^0)^{-1} \mathbf{b''} = \lambda^{-2} p_0^{-1} \mathbf{p''} =
- \lambda^{-2} \mathbf{y} =
\lambda^{-1} (\|\mathbf{p''}\|^2 + \lambda^{2} m^2)^{-1/2} \mathbf{p''},
\end{equation}
which show that the acceleration $\|\mathbf{a}\|$ has the upper bound $\lambda^{-1}$ and the distance $\|\mathbf{y}\|$ of the center of mass from the origin has the upper bound $\lambda$.  

The dinamical equation (\ref{PDot1}) follows from eq.\ (\ref{PDot4}) and the scalar constraint (\ref{Constr2}). Eq.\ (\ref{PDot3}) is automatically satisfied and from eqs. (\ref{PDot2}) and (\ref{PDot4}) we obtain
\begin{equation} \label{PDot5}
\dot\mathbf{p} = - p_0 \mathbf{b''} + b^0 \mathbf{f} 
- \mathbf{b'} \times \mathbf{p}, \qquad
\dot\mathbf{p}'' = - b^0 \mathbf{p}
- \mathbf{b}' \times \mathbf{p}''.
\end{equation}
There are no secondary constraints. 

As in the simple model considered in the preceding section, the time evolution of the quantities $b^0$, $\mathbf{b}'$ and $b^{\bullet}$ is not determined by the dynamical equations and the model has gauge invariances. Since there are five primary constraints and five arbitrary gauge variables, the phase space has dimension twelve.  We can impose the gauge fixing conditions $b^0 = 1$, $\mathbf{b}' = 0$ and we get the simpler dynamical equations
\begin{equation} \label{Exp}
\dot\mathbf{p} = - \lambda^{-2} \mathbf{p}'' + \mathbf{f}, \qquad 
\dot\mathbf{p}'' = - \mathbf{p}.
\end{equation}
This system is in normal form.

We obtain
\begin{equation} 
\frac{d}{d \tau} (p_0^2 -\|\mathbf{p}\|^2) = -2 \, \mathbf{f} \cdot \mathbf{p}.
\end{equation}
Even if initially the four-momentum is time-like, by applying a suitable force $\mathbf{f}$, it may become space-like. This means that one cannot simply discard as unphysical the states with space-like four-momentum.

We see from eqs.\ (\ref{Dyn6}), (\ref{EB}) and (\ref{EBf}) that, when the particle is subject to gravitational fields only, if $\mathbf{p}' = \mathbf{p}'' = 0$ we have $\mathbf{f} = 0$. It follows that $\mathbf{p}'' = \mathbf{b}'' = \mathbf{p} = 0$ is a solution which describes an ordinary spinless particle with mass $p_0 = m$ moving according to the laws of general relativity. We show below that, even in the absence of external fields, there are many other solutions, as it is expected, since the phase space has dimension larger than six.

It is shown in refs.\ \cite{NFS,NFS2} that, in space-times of constant curvature and in the absence of other external fields, the dynamical equations can be solved exactly. In fact, from eqs.\ (\ref{CCurv2}) and (\ref{Exp}) we obtain the equation
\begin{equation} 
\ddot\mathbf{p}'' = (\rho + \lambda^{-2}) \mathbf{p}''.
\end{equation}
For $\rho > - \lambda^{-2}$ it has exponential solutions and for $\rho < - \lambda^{-2}$ it has periodic solutions. Considering with more detail the case  $\rho = 0$, namely a flat spacetime, the general solution has the form
\begin{equation} 
\mathbf{p}'' = \mathbf{u} \exp(\lambda^{-1} \tau) + \mathbf{v} \exp(- \lambda^{-1} \tau).
\end{equation}

The acceleration $\|\mathbf{a}\|$ and the distance $\|\mathbf{y}\|$ of the center of mass from the origin of the moving frame are given by eq.\ (\ref{BB}). For generic initial conditions, they approach their maximum values when time increases or decreases, in particular
\begin{equation} 
\lim_{\tau \to \infty} \mathbf{a} = \lambda^{-1} \|\mathbf{u}\|^{-1} \mathbf{u}, 
\qquad \mathbf{u} \neq 0.
\end{equation}
For $\mathbf{u} = \mathbf{v} = 0$, we obtain $\mathbf{p}'' = \mathbf{a} = 0$, namely the uniform motion that is expected for an ordinary particle in the absence of external forces. This solution, however, is unstable, since small changes of the initial conditions lead to solutions with a large acceleration in the past or in the future. We have
\begin{equation} 
p_0^2 -\|\mathbf{p}\|^2 = m^2 + 4 \lambda^{-2} \mathbf{u} \cdot \mathbf{v}
\end{equation}
and we see that for some solutions the energy-momentum four-vector is spacelike. The accelerated solutions recall the run-away solutions of the Lorentz-Dirac equation \cite{Jackson}, with the important difference that in our case we are considering test particles and disregarding the radiated energy.

In a flat spacetime one can easily compute the motion of the origin of $s(\tau)$, namely of the point charge associated with the extended particle.  In the generic case, it approaches soon the velocity of light with respect to a fixed frame. Of course, the center of mass follows a linear world line and its distance from the charge becomes large in the fixed frame. In the moving frame, however, as an effect of the Lorentz transformation, it remains small. 

If, instead of considering a test particle, we take into account effects depending on higher powers of the charge $e$, it is possible that the radiated energy (bremsstrahlung) prevents the unstable behavior described above. In the following sections, however, we try a different solution of this problem.

\section{A model with a maximal \\ pseudo-acceleration.}

In order to avoid the unphysical energy-momentum spectrum and the unstable behaviour of the model described in the preceding section, a reasonable attempt is to consider zero-momentum frames instead of rest frames. This can be obtained by dropping the Lagrange multipliers from eq.\ (\ref{Lagr6}) namely by assuming
\begin{equation} 
L = - m \left((b^0)^2 - \lambda^2 \|\mathbf{b}''\|^2 \right)^{1/2} + e b^{\bullet}.
\end{equation}
Besides eq.\ (\ref{Carica}) we obtain the vector constraints
\begin{equation}  \label{VConstr}
\mathbf{p} = 0, \qquad \mathbf{p}' = 0,
\end{equation} 
showing that $s(\tau)$ is a zero-momentum frame and the particle is spinless. It is clear that the four-momentum $p$ is necessarily timelike. The equations (\ref{PP}), the scalar constraint (\ref{Constr2}) and eq.\ (\ref{BB}) of the preceding section are still valid and, in this case too, we consider only solutions with $p_0 \geq m$ and, as a consequence, $b^0 >0$. Eq.\ (\ref{BB}), however, does not give an upper bound to the acceleration, but to the pseudo-acceleration $\mathbf{a}$ defined in section VI. 

The relation
\begin{equation} 
p_0 = m \left(1 - \lambda^{-2} \|\mathbf{y}\|^2 \right)^{-1/2}
\end{equation}
shows that $p_0$ can be interpreted as a confining potential that binds the charge to the center of mass. When a force is applied to the charge, $\|\mathbf{y}\|$ and $p_0$ increase and this explains intuitively why the pseudo-acceleration remains bounded.

As in the model of the preceding section, the dynamical equation (\ref{PDot1}) follows from eq.\ (\ref{PDot4}) and the scalar constraint (\ref{Constr2}), eq.\ (\ref{PDot3}) is automatically satisfied and from eqs. (\ref{PDot2}) and (\ref{PDot4}) we obtain
\begin{equation} \label{DynE1}
p_0 \mathbf{b}'' = b^0 \mathbf{f}, \qquad
\dot \mathbf{p}'' = p_0 \mathbf{b} - \mathbf{b}' \times \mathbf{p}''.
\end{equation}

In this case too we have gauge symmetries. If we impose the gauge fixing conditions $b^0 = 1$ and $\mathbf{b}' = 0$, we get the simplified equations
\begin{equation} \label{DynE2}
\mathbf{p}''  = \lambda^2 p_0 \mathbf{b}'' = \lambda^2 \mathbf{f} 
= \lambda^2 (\hat \mathbf{E} + \mathbf{b} \times \hat\mathbf{B}), 
\end{equation}
\begin{equation} \label{DynE3}
\dot \mathbf{p}'' = p_0 \mathbf{b}.
\end{equation}
Note that the four-vector $(b^0, \mathbf{b})$, proportional to the four-velocity of the charge measured in the frame $s(\tau)$, is not necessarily timelike and in section X we find examples in which it is actually spacelike.  It follows that its time component in some other fixed reference frame $\hat s$ may vanish or even become negative. This means that the charge ``moves backwards in time", namely, more exactly, the time coordinate with respect to the frame $\hat s$ is a decreasing function of the parameter $\tau$. A further discussion will be given in section XIV.

As we have remarked in the preceding section, if there is only a gravitational field, but no other external fields, the equalities $\mathbf{p}' = \mathbf{p}'' = 0$ imply that $\mathbf{f} = 0$. It follows that the equalities $\mathbf{f} = \mathbf{p}'' = \mathbf{b}'' = \mathbf{b}' =\mathbf{b} = 0$ define solutions in which $s(\tau)$ is a parallel transported rest frame. These solutions describe an ordinary spinless particle with mass $p_0 = m$ moving according to the laws of general relativity, but in general there are other different solutions with $\mathbf{p}'' \neq 0$.  In a space of constant curvature $\rho$, as we see from eq.\ (\ref{CCurv2}), we have $\mathbf{f} = - \rho \mathbf{p}''$ and if $\rho \neq - \lambda^{-2}$, we have necessarily $\mathbf{f} = \mathbf{p}'' = 0$ and the only solutions are the ones described above. In this case, the phase space has dimension six.

If $\hat \mathbf{B} \neq 0$, from eq.\ (\ref{DynE2}) we obtain the secondary constraint
\begin{equation} \label{Proj}
(\hat \mathbf{E} -  \lambda^{-2} \mathbf{p}'') \cdot \hat\mathbf{B} = 0
\end{equation}
We have eight primary constraints, one secondary constraint and five arbitrary gauge variables. It follows that, if there are no tertiary constraints, the dimension of the phase space is eight. If $\hat\mathbf{B} = 0$ in the whole space $\mathcal{S}$, as in the constant curvature models treated above,  eq.\ (\ref{DynE2}) gives the three secondary constraints
\begin{equation} \label{SecCon}
\mathbf{p}'' 
= \lambda^2 \hat \mathbf{E}  
\end{equation}
 and the phase space has dimension six, as in the free case. 

\section{A particle in a constant electromagnetic field.}

It is useful to consider, as in section VII, a particle in a flat spacetime with a constant electromagnetic field. Some  results will be approximately valid for slowly varying fields. We consider first the case in which the invariants (\ref{Invar}) have the properties $I \geq 0, J = 0$ and in the frame $s(0)$ we have $\mathbf{B} = 0$.  If we choose the gauge $\mathbf{b}' = 0$, we see from eq.\ (\ref{VarBE}) and (\ref{DynE2}) that $\dot \mathbf{E} = \dot \mathbf{B} = 0$. It follows that along the whole trajectory $\mathbf{E}$, $\mathbf{p}''$ and $\mathbf{b}''$ are constant, $\mathbf{B} = 0$ and $\mathbf{b} = 0$. The frame $s(\tau)$ is a rest frame that moves with a constant acceleration given by
\begin{equation} \label{Accel}
\mathbf{b}'' =  (m^2 + \lambda^2  \|e\mathbf{E}\|^2)^{-1/2} e \mathbf{E}.
\end{equation}	
We see that, in this case, the singular manifold defined by $\mathbf{B} = 0$ is invariant, namely all the trajectories crossing it are completely contained in it. The acceleration remains bounded even if the electric field is very large. As in the case of section VII, $s(0)$ determines the initial conditions completely and the solution is stationary.

If in the frame $s(0)$ we have $\mathbf{B} \neq 0$, we must also specify two components of $\mathbf{p}''$ in order to describe the initial condition completely. We show, however, that there is a finite number of stationary solutions starting from $s(0)$ in which all the components of the field and all the quantities $b^{\alpha}$ and $p_{\alpha}$ are constant and $s(\tau)$ is given by the exponential formula (\ref{OPS}). The other, nonstationary, solutions will be treated perturbatively in the next section.

We assume first that $\mathbf{E} \times \mathbf{B} \neq 0$, since the case in which this vector vanishes needs a separate treatment.  If we require that the derivatives in eqs.\ (\ref{VarBE}) and (\ref{DynE1}) vanish, we obtain after some calculation,
\begin{eqnarray} \label{Stationary}
&\mathbf{b} = a \mathbf{E} \times \mathbf{B},& \nonumber \\ 
&\mathbf{b'} = p_0^{-1} e \left( - (1 - a \|\mathbf{B}\|^2) \mathbf{B} + 
a (\mathbf{B} \cdot \mathbf{E}) \mathbf{E} \right),& \nonumber \\ 
&\mathbf{b''} = p_0^{-1} e \left((1 - a \|\mathbf{B}\|^2) \mathbf{E} + 
a (\mathbf{B} \cdot \mathbf{E}) \mathbf{B} \right),& \nonumber \\ 
&p_0^2 = m^2 + \lambda^2 e^2 \left((1 - a \|\mathbf{B}\|^2)^2 \|\mathbf{E}\|^2
+ a (2 - a \|\mathbf{B}\|^2) (\mathbf{E} \cdot \mathbf{B})^2\right).&
\end{eqnarray}
The second eq.\ (\ref{DynE1}) is satisfied if
\begin{equation}
a p_0^2 = \lambda^2 e^2 \left((1 - a \|\mathbf{B}\|^2)^2 + a^2 (\mathbf{B} \cdot \mathbf{E})^2\right),
\end{equation}
namely, by substituting the expression for $p_0^2$,
\begin{eqnarray} \label{Alg}
&f(a) = (1 - a \|\mathbf{B}\|^2) \left((1 - a \|\mathbf{B}\|^2)
(1 - a \|\mathbf{E}\|^2) - a^2 (\mathbf{E} \cdot \mathbf{B})^2 \right) 
- \lambda^{-2} e^{-2} m^2 a& \nonumber \\
& = (1 - a \|\mathbf{B}\|^2) \left(1 - a \|\mathbf{B}\|^2 - a \|\mathbf{E}\|^2
+ a^2 \|\mathbf{E} \times \mathbf{B}\|^2 \right) - \lambda^{-2} e^{-2} m^2 a = 0. \quad &
\end{eqnarray}

This is, in general, a third degree algebraic equation in the variable $a$, that has one or three real solutions. The polynomial function $f(a)$ has the following properties
\begin{eqnarray} 
&f(a) > 0  \quad\mbox{for}\quad a \leq 0, \qquad f(0) = 1,& \nonumber \\
&f(a) < 0  \quad\mbox{for}\quad \|\mathbf{E}\|^{-2} \leq a \leq \|\mathbf{B}\|^{-2},& \nonumber \\
&f(a) < 0  \quad\mbox{for}\quad \|\mathbf{B}\|^{-2} \leq a \leq \|\mathbf{B}\|^{-2}
(1 + \lambda^{-2} e^{-2} m^2 \left(\|\mathbf{E}\|^2 + \|\mathbf{B}\|^2)^{-1}\right),& \nonumber \\
&f'(a) < - \lambda^{-2} e^{-2} m^2  \quad\mbox{for}\quad 0 \leq a \leq \min(\|\mathbf{E}\|^{-2}, \|\mathbf{B}\|^{-2}).&
\end{eqnarray}
It follows that $f(a)$ has only one zero in the interval 
\begin{equation} \label{Smaller}
0 < a < \min(\lambda^2 e^2 m^{-2}, \|\mathbf{E}\|^{-2}, \|\mathbf{B}\|^{-2})
\end{equation}
and, possibly, two zeros in the half line
\begin{equation} \label{Other}
a \geq \|\mathbf{B}\|^{-2}
\left(1 + \lambda^{-2} e^{-2} m^2 (\|\mathbf{E}\|^2 + \|\mathbf{B}\|^2)^{-1}\right).
\end{equation}

If $a$ lies in the interval, (\ref{Smaller}) we have $\|\mathbf{b}\|^2 < 1$, namely the velocity of the charge is smaller that the velocity of light. We see below that, for some values of the fields, if we choose a solution in the half line (\ref{Other}), the charge may move faster than light.  Note that if $\|\mathbf{b}\|^2 = \lambda^2 \|\mathbf{b'} \times \mathbf{b''}\|^2 > 1$, since $\lambda \|\mathbf{b''}\| < 1$, we have  $\lambda \|\mathbf{b'}\| > 1$, namely the frame $s(\tau)$ rotates, with all the dynamical vector variables,  with an angular velocity larger than $\lambda^{-1}$. If $\lambda$ is very small, one can only observe an average value of the velocity $\mathbf{b}$, that is negiglible, since $\mathbf{b}$ is orthogonal to the angular velocity $\mathbf{b'}$.

Note that the fields $\mathbf{E}$ and $\mathbf{B}$ are measured with respect to the moving frame $s(\tau)$ and if the particle is very fast can be much larger than the fields measured in the laboratory frame.  Even if we take this remark into account, if $\lambda$ is of the order of Planck's length, in all the experimental situations we have
\begin{equation} \label{Small}
\|\mathbf{E}\|, \, \|\mathbf{B}\| \ll \lambda^{-1} e^{-1} m.
\end{equation}
Under these conditions, the smallest solution of eq.\ (\ref{Alg}) is given by the approximate formula
\begin{equation} 
a_1 \approx \lambda^2 e^2 m^{-2}.
\end{equation}
Note that when $\lambda \to 0$, this solution tends to zero and the other two solutions either become complex or tend to infinity. In this limit we obtain, as it is expected, the ordinary model of section VII.

We have to consider the special case in which $\mathbf{B} \neq 0$ and $\mathbf{E} \times \mathbf{B} = 0$, namely $\mathbf{E} = k \mathbf{B}$. We obtain the following conditions for the stationary solutions
\begin{equation} \label{Stationary2}
\mathbf{b} = 0, \qquad \mathbf{b'} = h \|\mathbf{B}\|^{-1} \mathbf{B}, \qquad 
p_0 \mathbf{b''} = e \mathbf{E}, \qquad p_0^2 = m^2 + \lambda^2 \|e \mathbf{E}\|^2,
\end{equation}
where $h$ is an arbitrary constant that describes just a choice of the rotational gauge. Note that eq.\ (\ref{Accel}) is also valid in this more general situation.

It is interestig to see if these equations can be obtained as a limit of eqs.\ (\ref{Stationary}) for $\|\mathbf{E} \times \mathbf{B}\| \to 0$. The equation (\ref{Alg}) tends to the second degree equation
\begin{equation} 
(1 - a \|\mathbf{B}\|^2) (1 - a \|\mathbf{E}\|^2 - a \|\mathbf{B}\|^2) 
- \lambda^{-2} e^{-2} m^2 a = 0
\end{equation}
and one of the three solutions of eq.\ (\ref{Alg}) tends to infinity, more exactly we have 
\begin{equation} 
a_3 \approx (\|\mathbf{E}\|^2 + \|\mathbf{B}\|^2) \|\mathbf{E} \times \mathbf{B}\|^{-2}.
\end{equation}
It follows that for this solution we have $\|\mathbf{b}\| \to \infty$, showing that there are stationary solutions with the charge moving faster than light. The other two (necessarily positive) solutions are given by
\begin{equation} 
a_{1,2} = (2 \|\mathbf{B}\|^2(\|\mathbf{E}\|^2 + \|\mathbf{B}\|^2))^{-1}
(\|\mathbf{E}\|^2 + 2 \|\mathbf{B}\|^2 + \lambda^{-2} e^{-2} m^2 \pm \Delta^{1/2}).
\end{equation}
where
\begin{equation} 
\Delta = (\|\mathbf{E}\|^2 + \lambda^{-2} e^{-2} m^2)^2 
+ 4 \lambda^{-2} e^{-2} m^2 \|\mathbf{B}\|^2 > 0.
\end{equation}
The corresponding values of the parameter $h$ are
\begin{eqnarray} \label{Limits}
&h_{1,2} = p_0^{-1} e \|\mathbf{B}\| (a_{\pm} (\|\mathbf{E}\|^2 + \|\mathbf{B}\|^2) -1)& 
\nonumber \\
&= e (2p_0 \|\mathbf{B}\|)^{-1} (\|\mathbf{E}\|^2 + \lambda^{-2} e^{-2} m^2 \pm \Delta^{1/2}).&
\end{eqnarray}
Only for these values of the gauge parameter $h$ the solutions (\ref{Stationary2}) can be obtained as limits of the solutions (\ref{Stationary}).

\section{Perturbations and stability.}

The next step is to study perturbatively trajectories slightly different from the stationary ones described above. Since the infinitesimal perturbations satisfy linear differential equations with constant coefficients, we look for complex exponential solutions, namely we write
\begin{equation}
b^{\alpha} \to b^{\alpha} +
\Re \left(\delta b^{\alpha} \exp(z \tau)\right),
\end{equation}
where the constant quantities $z$ and $\delta b^{\alpha}$ are complex. A similar notation is used for the other variables.  The derivatives with respect to $\tau$ can be replaced by the factor $z$. We are not interested in perturbations with $z = 0$, which give other already known stationary solutions. For  $z \neq 0$ we obtain new nonstationary solutions, but also stationary solutions, modified by an infinitesimal time-dependent gauge transformation.

From eq.\ (\ref{VarBE}) we obtain
\begin{eqnarray} \label{Pert1}
&z \delta\mathbf{E} = \mathbf{b}'' \times \delta\mathbf{B} - \mathbf{B} \times \delta\mathbf{b}''
- \mathbf{b}' \times \delta\mathbf{E} + \mathbf{E} \times \delta\mathbf{b}',& \nonumber \\
&z \delta\mathbf{B} = - \mathbf{b}'' \times \delta\mathbf{E} + \mathbf{E} \times \delta\mathbf{b}''
- \mathbf{b}' \times \delta\mathbf{B} + \mathbf{B} \times \delta\mathbf{b}'.&
\end{eqnarray}
Since the invariats (\ref{Invar}) are not affected by the perturbation, we also require the relations
\begin{equation} \label{Pert2}
\mathbf{E} \cdot \delta \mathbf{E} = \mathbf{B} \cdot \delta \mathbf{B}, \qquad
\mathbf{E} \cdot \delta \mathbf{B} = - \mathbf{B} \cdot \delta \mathbf{E},
\end{equation}
which, however, are not independent from eqs.\ (\ref{Pert1}).

From the other dynamical equations we have
\begin{eqnarray} \label{Pert3}
&\delta\mathbf{p}'' = \lambda^2 e (\delta\mathbf{E} 
+ \mathbf{b} \times \delta\mathbf{B} - \mathbf{B} \times\delta\mathbf{b}),& \nonumber \\
&z \delta\mathbf{p}'' = p_0 \delta\mathbf{b} + \mathbf{b} \delta p_0
- \mathbf{b}' \times \delta\mathbf{p}'' + \mathbf{p}'' \times \delta\mathbf{b}',& \nonumber \\
&\delta\mathbf{p}'' = \lambda^2 \mathbf{b}'' \delta p_0 
+ \lambda^2 p_0 \delta \mathbf{b}'',& \nonumber \\ 
&\delta p_0 = \lambda^2 p_0^3 m^{-2} \mathbf{b}'' \cdot \delta\mathbf{b}''.& 
\end{eqnarray}
After the elimination of the variables $\delta p_0$, $\delta\mathbf{p}''$ and $\delta\mathbf{b}$, we obtain a system of nine homogeneous linear equations in the twelve unknown variables $\delta\mathbf{E}$, $\delta\mathbf{B}$, $\delta\mathbf{b}''$ and $\delta\mathbf{b}'$. 

For all the values of $z$ this system has three linearly independent solutions which represent rotational gauge transformations. They are given by
\begin{equation} 
\delta\mathbf{E} = \mathbf{E} \times \mathbf{r}, \qquad
\delta\mathbf{B} = \mathbf{B} \times \mathbf{r}, \qquad
\delta\mathbf{b}'' = \mathbf{b}'' \times \mathbf{r}, \qquad
\delta\mathbf{b}' = \mathbf{b}' \times \mathbf{r} + z \mathbf{r},
\end{equation}
where $\mathbf{r}$ is an arbitrary infinitesimal complex vector. Note that unless
\begin{equation} \label{Zeta}
z^2  = - \|\mathbf{b}'\|^2,
\end{equation}
the vector $\delta\mathbf{b'}$ can take arbitrary values and when we look for physically relevant perturbations we can choose the gauge $\delta\mathbf{b}'= 0$. We obtain in this way an homogeneous linear system of nine equation in nine unknowns, which has nonvanishing solutions only if its determinant $\det M(z)$ vanishes. If $z$ satisfies eq.\ (\ref{Zeta}), there is a pure gauge perturbation with $\delta\mathbf{b}'= 0$, and this implies that $\det M(\pm i \|\mathbf{b}'\|) = 0$. If we disregard these uninteresting solution, the other nonvanishing solutions of the algebraic equation $\det M(z) = 0$ correspond to physically relevant perturbations.

If  $\mathbf{E} \times \mathbf{B} \neq 0$, it is convenient to use the projections of all the  vectors on the basis formed by the vectors $\mathbf{E}$, $\mathbf{B}$ and $\mathbf{E} \times \mathbf{B}$. Eq.\ (\ref{Pert2}) permits the elimination of two unknowns and we have to calculate the determinant of a $7 \times 7$ matrix. It has almost thousand terms and one has to use a computer algebra program \cite{Vermaseren}. We find that the determinant $\det M(z)$ contains only the powers $z^6$, $z^4$ and $z^2$, and, since we know that it vanshes for $z^2 = - \|\mathbf{b}'\|^2$, we can find the other, physically relevant, solution of the eq.\ $\det M(z) = 0$ by means of Ruffini's rule. With an appropriate choice of the variables, the computer calculation gives the unexpectedly simple result
\begin{equation} \label{ZetaPh}
- z^2 = \omega^2 = \lambda^{-4} p_0^2 \|e \mathbf{B}\|^{-2} + 4 \lambda^{-2} m^2 p_0^{-2} > 0,
\end{equation}
where $p_0^2$ is given by eq.\ (\ref{Stationary}). We see that $z = \pm i \omega$ is pure imaginary and this means that the stationary solutions are stable with respect to linear perturbations. The perturbations have a frequency $\omega / 2\pi$, that when $\|e \mathbf{B}\| \to 0$,  tends to infinity explaining why a degree of freedom disappears for  $\|e \mathbf{B}\| = 0$. If we use eq.\ (\ref{Small}), we see that the frequency $\omega / 2\pi$ is much larger than the Planck frequency $\lambda^{-1}$. 

In order to extend our results to the case in which $\mathbf{E} \times \mathbf{B} = 0$ a separate discussion, that does not require computer algebra, is needed. From eqs.\ (\ref{Stationary2}) (\ref{Pert2}) and (\ref{Pert3}) we find 
\begin{equation} 
\mathbf{B} \cdot \delta \mathbf{E} = \mathbf{B} \cdot \delta \mathbf{B} 
= \mathbf{B} \cdot \delta \mathbf{p''} =  \mathbf{B} \cdot \delta \mathbf{b} = 0, \qquad
\delta p_0 = 0.
\end{equation}
As in the general case, we choose $\delta \mathbf{b'} = 0$.

After some calculations we obtain the relation
\begin{equation} 
z (\delta \mathbf{B} - \lambda^2 k e p_0^{-1} \|\mathbf{B}\|^2 \delta \mathbf{p''}) 
= - \mathbf{b'} \times (\delta \mathbf{B} 
- \lambda^2 k e p_0^{-1} \|\mathbf{B}\|^2 \delta \mathbf{p''})
\end{equation}
and if $z^2 \neq - \|\mathbf{b'}\|^2$ we have
\begin{equation} 
\delta \mathbf{p''} = \lambda^{-2} (k e)^{-1} p_0 \|\mathbf{B}\|^{-2} \delta \mathbf{B}.
\end{equation}

By substituting this formula and eq.\ (\ref{Stationary2}) into eq.\ (\ref{Pert1}), we obtain a linear homogeneous system of the form
\begin{equation} 
z \left( \begin{array}{cc}\delta \mathbf{E} \\ \delta \mathbf{B}\end{array} \right)
= \|\mathbf{B}\|^{-1}  A
\left( \begin{array}{cc}\mathbf{B} \times \delta \mathbf{E} \\ \mathbf{B} \times \delta \mathbf{B}\end{array} \right),
\end{equation}
where the matrix $A$ is given by
\begin{equation} 
A = \left( \begin{array}{cc} - h & \lambda^{-2} m^2 p_0^{-1} \|\mathbf{eE}\|^{-1} \\
- p_0^{-1} \|\mathbf{eE}\| & -h + \lambda^{-2} p_0 \|\mathbf{eB}\|^{-1} \end{array} \right),
\end{equation}
Since the vectors $\delta\mathbf{E} $ and $\delta\mathbf{B}$ must be perpendicular to $\mathbf{B}$, by iterating this formula we obtain
\begin{equation} 
z^2 \left( \begin{array}{cc}\delta \mathbf{E} \\ \delta \mathbf{B}\end{array} \right)
= - A^2 \left( \begin{array}{cc}\delta \mathbf{E} \\ \delta \mathbf{B}\end{array} \right).
\end{equation}

One can easily show that this linear system has nonvanishing solutions if
\begin{equation} 
z^2 = - (h_{1,2} - h)^2,
\end{equation}
where the quantities $h_{1,2}$ are given by eq.\ (\ref{Limits}). From the same equation and eq.\ (\ref{ZetaPh}) we obtain the identity
\begin{equation} 
h_2 - h_1 = \omega.
\end{equation}

We see that, in this case too, the stationary trajectories are stable, but the frequency of the perturbations depends on the gauge parameter $h$. There are two independent perturbations with different angular velocities and only their difference is invariant under the rotational gauge and physically relevant. If we choose $h = h_{1,2}$, in both cases we obtain the solution $z^2 = 0$ and the solutions given by eq.\ (\ref{ZetaPh}), namely the limits of the solutions found for $\mathbf{E} \times \mathbf{B} \neq 0$.

\section{A particle in a simple gravitational field.}

In order to understand the properties of the model in a gravitational field, it is instructive to consider, besides the spacetimes with constant curvature we have already examined in section IX, another simple model, namely the stationary Einstein cosmological model \cite{MTW} with spatial curvature $\rho$. The curvature tensor in the frame $s(\tau)$ takes the form
\begin{eqnarray} \label{RW}
&F_{i k}^{[j l]} = \rho (\delta_i^j \delta_k^l - \delta_k^j \delta_i^l - \delta_i^j v_k v^l + \delta_k^j v_i v^l + \delta_i^l v_k v^j - \delta_k^l v_i v^j),& \nonumber \\
&v_k v^k = 1, \qquad \rho > 0,
\end{eqnarray}
where, in a cosmological interpretation, $v= (v_0, \mathbf{v})$ is the four-velocity with respect to the frame $s \in \mathcal{S}$ of a privileged frame in which the matter and radiation distribution  is isotropic. 

From eq.\ (\ref{Dyn6}) we obtain
\begin{equation}
G_{ik} =  - \rho (p_{[ik]} + v^j p_{[ji]} v_k - v^j p_{[jk]} v_i),
\end{equation}
or, in the vector notation, 
\begin{equation}
\hat \mathbf{E} = \rho (\|\mathbf{v}\|^2 \mathbf{p''} 
- (\mathbf{v} \cdot \mathbf{p''}) \mathbf{v}), \qquad 
\hat \mathbf{B} =  \rho v_0 \, \mathbf{v} \times \mathbf{p''}.
\end{equation}

Eqs.\ (\ref{DynE2}) (\ref{DynE3}) give the result
\begin{equation} 
(\lambda^{-2} \rho^{-1} - \|\mathbf{v}\|^2) \mathbf{p''} 
+ (\mathbf{v} \cdot \mathbf{p''}) \mathbf{v}
= p_0^{-1} v_0 \dot\mathbf{p}'' \times (\mathbf{v} \times \mathbf{p''}),
\end{equation}
and the constraint (\ref{Proj}) is identically satisfied. We assume that $\|\mathbf{v}\|^2 < \lambda^{-2} \rho^{-1}$. If $\mathbf{v} \times \mathbf{p''} = 0$ we must have $\mathbf{p''} = 0$, otherwise we obtain
\begin{equation} \label{Motion}
\mathbf{v} \cdot \dot\mathbf{p}'' = - p_0 v_0^{-1} (\lambda^{-2} \rho^{-1} - \|\mathbf{v}\|^2), \qquad
\dot p_0 = \lambda^{-2} v_0^{-1} (\mathbf{v} \cdot \mathbf{p''}).
\end{equation}
The component of $\dot\mathbf{p}''$ normal to the vectors $\mathbf{v}$ and $\mathbf{p''}$ is not determined by the dynamical equations, namely there is a gauge symmetry. We fix the gauge by means of the condition
\begin{equation} \label{Fix}
\dot\mathbf{p}'' \cdot \mathbf{v} \times \mathbf{p''} = 0.
\end{equation}

One can show that the covariant spacetime derivatives $A_i v^k$ vanish and it follows that the derivatives of $v^k$ with respect to the parameter $\tau$ are given only by the boost of the Lorentz frame, namely 
\begin{equation} \label{Approx}
\dot\mathbf{v} = - v_0 \mathbf{b''} = - \lambda^{-2} p_0^{-1} v_0 \mathbf{p''}, \qquad
\dot v_0 = - (\mathbf{v} \cdot \mathbf{b''})
= - \lambda^{-2} p_0^{-1} (\mathbf{v} \cdot \mathbf{p''}).
\end{equation}

From the preceding equations it follows
\begin{equation} \label{Ineq}
0 \leq \|\mathbf{v} \times \mathbf{p''}\|^2 = \lambda^2 (p_0^2 - m^2) \|\mathbf{v}\|^2
- \lambda^4 v_0^2 (\dot p_0)^2,
\end{equation}
\begin{equation} 
\frac{d w}{d \tau} = 0, \qquad w = p_0 v_0 \geq m.
\end{equation}
The conserved quantity $w$ is the energy measured in a privileged frame. We also obtain
\begin{equation} \label{SecDer}
\frac{d^2}{d \tau^2} \log p_0 = - \lambda^{-2} v_0^{-2} (\lambda^{-2} \rho^{-1}  + 1 
- w^2 m^2 p_0^{-4}).
\end{equation}
If $w$ is not extremely large, the right hand side is negative, $p_0(\tau)$ has a maximum for $\tau = \tilde \tau$ and violates the inequality $p_0 \geq m$ outside a given interval. A stronger limitation is given by eq.\ (\ref{Ineq}).  It is clear that $\tau$ cannot vary in the whole real line.

For a detailed discussion, it is convenient to introduce some simplifications. We disregard higher powers of the adimensional quantity $\lambda^2 \rho_0$, which is extremely small in all the interesting cases and we also disregard higher powers of $\tau - \tilde\tau$. We put $p_0(\tilde\tau) = \tilde p_0$, $v_0(\tilde\tau) = \tilde v_0$, $\mathbf{v}(\tilde\tau) = \tilde\mathbf{v}$ and we obtain the approximate solution
\begin{equation} 
p_0(\tau) \approx \tilde p_0 (1 - 2^{-1} \lambda^{-4} \rho^{-1} \tilde v_0^{-2} 
(\tau - \tilde\tau)^2),
\end{equation}
\begin{equation} 
\dot p_0(\tau) \approx  - \tilde p_0 \lambda^{-4} \rho^{-1} \tilde v_0^{-2} (\tau - \tilde\tau).
\end{equation}

From eq.\ (\ref{Ineq}) we see that $\tau$ must belong to the interval
\begin{equation} \label{Interval}
|\tau - \tilde\tau| < 2^{-1} \Delta\tau \approx (1 - m^2 \tilde p_0^{-2})^{1/2} \lambda^3 \rho \tilde v_0 \|\tilde\mathbf{v}\|.
\end{equation} 
This inequality shows that it was consistent to disregard higher powers of $\tau - \tau_0$ together with higher powers of $\lambda^2 \rho$. In this interval we have
\begin{equation} 
0 \leq \tilde p_0 - p_0(\tau) \leq  
2^{-1} \lambda^2 \rho \tilde p_0 (1 - m^2 \tilde p_0^{-2}) \|\tilde\mathbf{v}\|^2
\end{equation}
and the quantities $p_0$, $\|\mathbf{p''}\|$, $v_0$, and $\mathbf{v}$ can be considered as constant. 

It follows from eq.\ (\ref{Ineq}) that at the end points of the interval (\ref{Interval}) $\mathbf{p''}$ is parallel to $\mathbf{v}$ and by taking eq.\ (\ref{Fix}) into account, we see that the vector $\mathbf{p''}$ moves approximately on an half circumference in a plane containig the vector $\mathbf{v}$. From eq.\ (\ref{Motion}) we see that the projection of $\dot\mathbf{p}''$ in the direction of $\mathbf{v}$ is constant and it follows that the component of $\dot\mathbf{p}''$ normal to $\mathbf{v}$ tends to infinity at the end points. 

The connection between $\tau$ and the ``cosmic'' time $t$, measured in the privileged rest frames defined by $\mathbf{v} = 0$, is 
\begin{equation} 
\frac{d t}{d \tau} = v_i b^i = v_0 - \mathbf{v} \cdot \mathbf{b}
= (\lambda^{-2} \rho^{-1} + 1) v_0^{-1} 
\end{equation}
and it follows that the interval (\ref{Interval}) measured in terms of the time $t$ is
\begin{equation} 
\Delta t \approx 2 \lambda (1 - m^2 p_0^{-2})^{1/2} \|\mathbf{v}\|,
\end{equation}
namely it is, in general, very small. Note that it does not depend on the curvature $\rho$, but only on the four-vector $v$ that defines the privileged frames.

The singularities at the end points of the interval (\ref{Interval}) can be avoided by using, instead of the parameter $\tau$ defined by the condition $b^0 = 1$, another parameter and we can also try to extend the solution outside the interval. This is possible only if, generalizing our Lagrangian formalism, we accept negative values of $b^0$, namely negative values of the square root in eq.\ (\ref{PP}).  We do not try to give in this paper a physical meaning to this extension, for instance in terms of antiparticles.

In any case, we have shown that, in the presence of an arbitrarily small gravitational field, the geodisic world lines, corresponding to the initial condition $\mathbf{p''} = 0$, are badly unstable, since an arbitrarily small change of $\mathbf{p''}$ leads to completely different, physicaly unacceptable, solutions.

\section{Another model.}

In the preceding sections we have examined two models that imply an upper bound to acceleration or to pseudoacceleration. However, these models have unwanted properties, in particular the energy-momentum  or the four-velocity of the charge may become spacelike and unstable solutions may appear. In order to show that it is not easy to avoid all these problems, we consider a model containing a fundamental length, in which both the energy-momentum and the four-velocity are necessarily timelike, as in the ordinary model of section VII. A simple and natural Lagrangian with these properties is
\begin{equation} 
L = - m \left((b^0)^2 - \|\mathbf{b}\|^2 - \lambda^2 \|\mathbf{b}''\|^2 \right)^{1/2} + e b^{\bullet}.
\end{equation}

The canonical momenta are given by
\begin{equation} 
\mathbf{p}' = 0, \qquad p_{\bullet} = - e,
\end{equation} 
\begin{eqnarray} 
&p_0 = m b^0 \left((b^0)^2 - \|\mathbf{b}\|^2 - \lambda^2 \|\mathbf{b}''\|^2 \right)^{-1/2},& \nonumber \\
&\mathbf{p} = m \left((b^0)^2 - \|\mathbf{b}\|^2 - \lambda^2 \|\mathbf{b}''\|^2 
\right)^{-1/2} \mathbf{b}&
\nonumber \\
&\mathbf{p}'' = m \lambda^2 \left((b^0)^2 - \|\mathbf{b}\|^2 - \lambda^2 \|\mathbf{b}''\|^2 
\right)^{-1/2} \mathbf{b}''&
\end{eqnarray}
and satisfy the scalar primary constraint
\begin{equation}  \label{Constr3}
(p_0)^2 - \|\mathbf{p}\|^2 - \lambda^{-2} \|\mathbf{p}''\|^2 = m^2,
\end{equation}
which assures that the energy-momentum is timelike.

We adopt the the gauge fixing conditions $b^0 = 1$ and $\mathbf{b}' = 0$ and we obtain the formulas
\begin{equation} 
\mathbf{b} = p_0^{-1} \mathbf{p} =
(\lambda^{-2} \|\mathbf{p}''\|^2 + \|\mathbf{p}\|^2 +m^2)^{-1/2} \mathbf{p},
\end{equation}
\begin{equation} 
\mathbf{b}'' = \lambda^{-2} p_0^{-1} \mathbf{p''} 
= \lambda^{-1} (\|\mathbf{p}''\|^2 
+ \lambda^{2} \|\mathbf{p}\|^2 + \lambda^{2} m^2)^{-1/2} \mathbf{p}'',
\end{equation}
which assure that $\|\mathbf{b}\| < 1$ and $\|\mathbf{b''}\| < \lambda^{-1}$. Note, however, that  $s(\tau)$ is neither a rest frame nor a zero-momentum frame and $\mathbf{b''}$ cannot be interpreted as the acceleration or the pseudoacceleration of the particle.

The dinamical equation (\ref{PDot1}) follows from eq.\ (\ref{PDot4}) and the scalar constraint (\ref{Constr3}). Eq.\ (\ref{PDot3}) is automatically satisfied and from eqs. (\ref{PDot2}) and (\ref{PDot4}) we obtain,  
\begin{equation} 
\dot{\mathbf{p}} = - p_0 \mathbf{b}'' + \mathbf{f} = - \lambda^{-2} \mathbf{p}'' + \mathbf{f}, \qquad
\dot{\mathbf{p}}'' = p_0 \mathbf{b} - \mathbf{p} = 0.
\end{equation}
We see that $\mathbf{p}''$ is conserved. If it vanishes, the dynamical equations coincide exactly with the equations of the ordinary model of section VII in its last version and there is no upper bound to the acceleration or to the pseudoacceleration.

\section{Conclusions.}

We have developed a general Lagrangian formalism for the description of test particles in gravitational and electromagnetic fields. It is based on a moving Lorentz frame associated to the particle and it is useful for the introduction of Lagrangians that depend on a fundamental length $\lambda$ and to discuss a possible upper bound to the proper acceleration. Other interesting effects may be suggested by a detailed discussion of suitable models.

The test particles introduced in this way present some features characteristic of extended objects, in particular, additional degrees of freedom are present, the charge, assumed to be concentrated at the origin of the moving frame, does not necessarily coincide with the center of mass, the energy-momentum four-vector is not necessarily parallel to the four-velocity and the definition of the acceleration is somehow ambiguous.

We have considered with more detail three Lagrangians depending on the acceleration of the frame, but not on its angular velocity, discussing, in particular, the following requirements:
\begin{itemize}
\item The energy-momentum four-vector lies in the future cone;
\item The velocity of the point charge is smaller than the velocity of light, namely the four-velocity is timelike;
\item The solutions have good stability properties;
\item There is an upper bound to the acceleration or to some related quantity as the pseudo-acceleration defined in section VI.
\end{itemize}

We have not been able to find a model that satisfies all these requirements and it is reasonable to conjecture that such a model does not exist within the formalism we have considered. Perhaps one should introduce new degrees of freedom that have not a geometric interpretation in terms of a moving frame. 

The model deascribed in section VIII, based on rest frames, has an unphysical energy-momentum spectrum and has unstable run-away solutions even in the absence of external fields.  

The model introduced in section IX, based on the zero-momentum frames, has unstable solutions when a nonconstant curvature is present and, under some circumstances, the velocity of the charge may become arbitrarily large. Nevertheless, it has some interesting properties. It always has a timelike energy-momentum four-vector and it has good stability properties if only a slowly varying electromagnetic field is present. 

An electric charge faster than light may be embarrassing, but does not seem to contradict the general principles of Maxwell's theory. The charge describes a singular distribution inside the particle of the electric current four-vector, that is not required to be time-like. Moreover, in the absence of curvature, the velocity vector rotates very fast and one can only observe its average value. 

In a constant electromagnetic field the second model has ``stationary'' solutions,  similar to the trajectories of an ordinary point particle with some small corrections, but, since the model has additional degrees of freedom, more complicated solutions exist, obtained by adding to a stationary solution a periodic perturbation with a frequency $\omega / 2\pi$  much larger than $\lambda^{-1}$. 

In a quantized version of the model, according to Bohr's correspondence principle, this frequency  is approximately interpreted as the frequency of the radiation emitted when an excited state of the particle decays. Since the frequency is defined in the zero momentum frame, the energy  $\hbar \omega$ has to be interpreted as the additional mass of the excited state. This mass is very large and in ordinary conditions there is not enough energy to excite the periodic degrees of freedom and they are ``frozen'', as it happens, for instance, to the nuclear degrees of freedom in low-energy atomic physics. These features of the quantized model may provide some very simplified insight into the connection between ordinary and ``transplanckian'' physics.

The model presented in section XIII has no problem with the properties of the energy-momentum and of the four-velocity, but the presence of the fundamental length $\lambda$  in the Lagrangian does not impose any limitation to the acceleration.

Of course, a more precise physical discussion requires a formal quantization procedure of the models, based on Dirac's treatment of constrained systems. In particular, the ground state energy corresponding to the additional (internal) degrees of freedom, which cannot be evaluated by means of the correspondence principle, could give an unacceptably large contribution to the mass. The treatment of many particle states should be given in terms of free quantum fields. For the model of section VIII these problems are treated in ref.\ \cite{NFLS} and we shall discuss some other models elsewhere.

\end{document}